# Coupling two distant double quantum dots to a microwave resonator

Guang-Wei Deng[1,2,*], Da Wei[1,2,*], Shu-Xiao Li[1,2], J. R. Johansson[3], Wei-Cheng Kong[1,2], Hai-Ou Li[1,2], Gang Cao[1,2], Ming Xiao[1,2], Guang-Can Guo[1,2], Franco Nori[4,5], Hong-Wen Jiang[6] and Guo-Ping Guo[1,2,+]

[1] Key Laboratory of Quantum Information, University of Science and Technology of China, Chinese Academy of Sciences, Hefei 230026, China

[2] Synergetic Innovation Center of Quantum Information and Quantum Physics, University of Science and Technology of China, Hefei, Anhui 230026, China

[3] iTHES research group, RIKEN, Wako-shi, Saitama, 351-0198 Japan

[4] CEMS, RIKEN, Wako-shi, Saitama, 351-0198 Japan

[5] Physics Department, The University of Michigan, Ann Arbor, Michigan 48109-1040, USA

[6] Department of Physics and Astronomy, University of California at Los Angeles, California 90095, USA

* These authors contributed equally to this work.

+ Correspondence to: gpguo@ustc.edu.cn

# Abstract

With recent advances in the circuit quantum electrodynamics (cQED) architecture[1,2], hybrid systems that couple nano-devices to microwave resonators have been developing rapidly[3-7]. Here we report an experimental demonstration of two graphene double quantum dots (DQDs) coupled over a distance of up to 60 μm, through a microwave resonator. We jointly measure the two DQDs' coupling to the resonator, which causes a nonlinear response in the resonator reflection amplitude in the vicinity of the degeneracy points of the two DQDs. This phenomenon is explained by the Tavis-Cummings (T-C) model[8]. We further characterize this nonlocal coupling by measuring the correlation between the DC currents in the two DQDs. This correlation is observed to be strongly dependent on the average photon number in the resonator. Our results explore T-C physics[8] in electronic transport, and also contribute to the study of non-local transport and future implementations of remote electronic entanglement[9-12].

# Introduction

The interaction between atoms and photons has been widely studied in cavity QED[13], and circuit QED has extended this idea to on-chip superconducting qubits[1,2,14]. Recent theoretical[9-12,15-19] and experimental[3-7,20] studies have also implemented this architecture with quantum dots by coupling them to resonators. So far, resonators have been coupled to quantum dots made of GaAs[4,6], carbon nanotubes[3,21], InAs nanowires[5] and graphene[20]. Beyond that, photon-mediated distant coupling between two single quantum dots (SQDs) has been studied[7]. Recently, in theoretical work on DQDs interacting through resonators, it has been proposed[9-12] that this setup can be used to entangle macroscopically-separated electron transport, which has applications in nanoscale quantum information processing and Bell inequality tests. A first step towards these goals would be an experimental demonstration of photon-mediated nonlocal electronic transport effects in separated mesoscopic quantum systems.

In general, the energy-level splitting in a DQD is easier to tune than in a SQD[26]. In a DQD, the energy splitting can be directly controlled through the gate-induced detuning, and can be tuned to an energy scale that is close to that of the resonator photons[26]. Motivated by this, we here use DQDs to investigate the dispersive DQD-resonator coupling near the charge-degeneracy points of the two DQDs. We report an experimental demonstration of coupling, through a microwave resonator, between two distant DQDs which are separated by about 60 μm. When sweeping the detuning of each DQD, in the proximity of the charge-degeneracy points, a dip is observed in the resonator reflection amplitude due to nonadditive dispersive contributions from the two DQDs. This phenomenon is explained by the Tavis-Cummings model[8], and it demonstrates the simultaneous dispersive coupling between one photonic mode and two DQDs. Moreover, with finite-bias voltages, the current through one of the DQDs is affected by the current through the other. By changing the microwave power applied to the resonator, this interaction can be controlled. This correlation between currents is studied with one DQD dispersively coupled to the resonator, while the coupling between the other DQD and the resonator can be tuned from dispersive to resonant.

## Results

**Device characterization.** Our sample is mounted in a dry dilution refrigerator, with a base temperature of about 38 mK. The resonator has a fundamental frequency $f_0$ of about 6.35086 GHz and a quality factor of about 3100. The hybrid device is shown in Fig. 1a,b,c (see Methods). Two etched graphene DQDs[22,23] made of separated (about 60 μm) few-layer flakes, are coupled to a superconducting reflection-line resonator[20,24,25] (RLR) through their sources. Transport measurements are performed for each DQD (see Fig. 2c,e). From the obtained honeycomb and Coulomb diamond diagrams, we characterize the DQDs by their charging energies and lever arms[26]. Meanwhile, charge-stability diagrams of both DQDs can also be obtained via the dispersive readout of the resonator (Fig. 2d,f). Using the method described in Ref. 20, we further characterize the device. The DQD-resonator coupling strength $g_{Ci}$, the tunnel coupling strength $2t_{Ci}$ and dephasing rate $\gamma_{2i}$ for the $i$th DQD are obtained[26].

**Coupling two graphene DQDs to a resonator.** We first confirm that the direct capacitive coupling between the two DQDs is negligibly small[26]. By tuning each DQD simultaneously across the SQD charging lines of each DQD[3,7,27] (this process corresponds to the adding or removing of one electron into or from the dot) and measuring the charging energy levels of both QDs using the resonator signal, we can extract the slopes of the SQD charging lines versus the gate voltages. We find that the charging lines are nearly horizontal or vertical, suggesting negligible capacitive coupling[26] (see Fig. S3e,f).

Compared to the SQD charging energy, $E_c \approx 2$ meV, the energy scale $2t_C$ is much closer to the resonator photon energy, $hf_0 \approx 27$ μeV. Thus, we expect that the DQDs can interact via the resonator, when both DQDs have $2t_C$ comparable to $hf_0$ and are operated near their charge-degeneracy points. First, we tune the two DQDs to the proximity of interdot charge transition lines that correspond to near-6-GHz tunnel coupling. Next, we sweep the detunings (each along the dashed lines, shown in Fig. 3a,b) and record the resonator signal. Figure 3d,e show the experimental results under a 0.10 pW (-100 dBm) applied microwave power. Near the center (corresponding to the charge-degeneracy points of both DQDs) the reflection amplitude is distinctly different from other regions, in that the contributions of the two DQDs are nonadditive. This observation matches well with the results of the T-C model with the Hamiltonian:

$$H = \omega_0 a^\dagger a + \sum_{i=1,2}[\frac{1}{2}\Omega_i \sigma_{zi} + g_i(\sigma_{+i} a + \sigma_{-i} a^\dagger)],$$

where $g_i = g_{Ci}\frac{2t_{Ci}}{\Omega_i}$, $\Omega_i = \sqrt{(2t_{Ci})^2 + \varepsilon_i^2}$. Here $\omega_0$ is the resonant frequency of the resonator and $\varepsilon_i$ denotes the detuning of DQD*i*. This model describes two two-level systems that are coupled to a photonic field. Using this model with the obtained parameters, we can reproduce the experimental amplitude and phase diagram[26] (Fig. 3f,g). The tunnel coupling strengths for both DQDs are 7.2 GHz[26], and the DQDs are therefore in the dispersive regime ($\Omega_i - hf_0 \gg g_i$).

We can understand this phenomenon as follows. Since the two DQDs are coupled to the resonator, they can both cause frequency shifts. Particularly, when the DQDs have zero detuning, they both significantly contribute to the dispersive interaction. These contributions add linearly, however, the amplitude and phase shifts of the resonator response are non-linear. In other words, as shown figure 3d, instead of reaching a larger amplitude shift (of greater absolute value), the cross-center region have a far lower one, as if the two shifts compete with and cancel out each other. This is a natural result of the T-C model and can be reproduced in simulations[26]. However, limited by the large dephasing rates in DQD systems, vacuum Rabi splitting and energy anti-crossings[28] have not been observed in our device, restricting us from further exploring its quantum information applications. However, this DQD-resonator system described by the Tavis-Cummings model leaves us with opportunities to study interesting aspects of nonlocal electronic transport properties[9-11].

**Photon-mediated electron transport.** Several theoretical works have recently predicted photon-mediated electron transport in DQDs-resonator hybrid systems[9-11]. Inspired by these predictions, we repeated the gate-sweeping procedures of the joint readout, but with focus on the DC current signals $I_{\text{DQD1(2)}}$ instead. Unless stated otherwise, the bias voltage is 60 $\mu V$ for both DQDs (schematically shown in Fig. 5a) throughout this part of the experiment. The DQDs are tuned to sites where $2t_{C1} > hf_0$ while $2t_{C2} \approx 6.1 \text{ GHz} < hf_0$. Figure 4a shows $I_{\text{DQD1}}$ and $I_{\text{DQD2}}$ as a function of $\varepsilon_1$ and $\varepsilon_2$. $I_{\text{DQD1}}$ decreases the most when $\varepsilon_2 = -2 \text{ GHz}$, where DQD2 is in resonance with the resonator photon, i.e., $hf_0 = \Omega_2 = \sqrt{\varepsilon_2{}^2 + (2t_{C2})^2}$. Fixing $\varepsilon_1$ at zero, we sweep $\varepsilon_2$ (shown as the horizontal white dashed line in Fig. 4a) under a series of microwave powers. The result indicates that $I_{\text{DQD1}}$ is influenced by DQD2 (Fig. 4b). Furthermore, if we view DQD2 as a switch whose on and off states denote whether DQD2 is on resonance ($\Omega_2 \sim hf_0$) or off resonance ($\Omega_2 \gg hf_0$) with the resonator, such a switch is able to control the resonator photonic field strength. From Fig. 4b one can extract the difference in $I_{\text{DQD1}}$, $\delta I$, for when DQD2 is on and off resonance. By converting $\delta I$ into a difference in the average photon number $\delta N$, we can study DQD2's effect on the photonic field strength in the resonator (i.e., the average photon number N). To this end, we first employ the empirical law $I_{\text{DQD1}}^{\varepsilon_1=0} = 32.73/(1 + 5P^{-2})$ as shown in figure 4c, with the current in units of pA and power in pW. 1 pW (-90 dBm) of applied power corresponds to about $5 \times 10^4$ photons in the resonator of our device[3,26,29], when both DQDs are in the Coulomb blockade regimes. Thus, when DQD2 is in the blockade regime, $I_{\text{DQD1}}^{\varepsilon_1=0}$ as a function of N can be approximated as $I_{\text{DQD1}}^{\varepsilon_1=0} = 32.73/(1 + 2 \times 10^{-9} N^{-2})$. The microwave response of the current through quantum dots has been studied previously, theoretically via the quantum photovoltaic effect in DQDs[18], experimentally in the microwave response of a SQD made of GaAs[29], and in the photon-induced current in graphene QDs at visible wavelengths[30]. The current-power relation we obtained above (see Fig. 4c) can be explained as electron heating by the resonator microwave field[29]. As Fig. 4d indicates, we find that $\delta N$ depends linearly on $P$ and thus on $N$. A physical consequence of this linear relation is that, in this power range, the total photon number in the resonator changes by a constant factor (~36%) when DQD2 is on and off resonance[26].

One can understand this nonlocal interaction as follows: When the energy level of DQD2 is near resonance ($\Omega_2 \sim hf_0$), and $\varepsilon_2 < 0$, it strongly absorbs photons from the resonator[31,32] (Fig. 5b), weakening the photonic field in the resonator and the microwave heating effect on DQD1. In other words, when DQD2 is near resonance, it leads to stronger dissipation for resonator photons[31]. Though DQD2 may also emit photons near the resonance condition and $\varepsilon_2 > 0$, the emission efficiency is too small[32] to be observed even for the largest current in our device. For these reasons, $I_{\text{DQD1}}$ ($R_{\text{DQD1}}$) only shows a dip (peak) as a function of $\varepsilon_2$ (Fig. 4b, 5c). For instance, the resistance $R_{\text{DQD1}}$ can be tuned by $\varepsilon_2$ from about 300 to 500 M$\Omega$ under 0.20 pW power. Finally we fix the gate condition for DQD1 and tune DQD2 from the blockade regime to its current peak center. $I_{\text{DQD1}}$ is then found to decrease nearly linearly with respect to $I_{\text{DQD2}}$. In this sense, they show a negative correlation and we establish a nonlocal control mediated by resonator photons (Fig. 5d).

## Discussion

Though there is still a long way to go before reaching the strong-coupling regime (coupling strength larger than the decoherence rates) in a DQD-resonator hybrid system, the large coupling strength (tens of MHz) opens up the possibility to study the interaction between two distant qubits made of quantum dot circuits. Compared to previous work[7] on SQDs, our DQD-based devices offer tunable two-level systems with energy scales closer to the resonator resonance, making it easier to reach the photon-DQD resonance condition. In the dispersive regime, we have observed a dip in the reflected amplitude, described by the T-C model, near the point where $\varepsilon_{1,2} = 0$ (Fig. 3d,f). Moreover, when one DQD satisfies the condition $hf_0 = \Omega = \sqrt{\varepsilon^2 + (2t_\text{C})^2}$, it can strongly affect the microwave field in the resonator, which in turn affects the other DQD. This distant interaction is activated by the microwave signal applied to the resonator.

In the analysis of photon-mediated transport, we estimate the attenuation throughout our measurement setup to be -75dB. However, this value differs from sample to sample with a standard deviation of ~3dB. Thus, while the fitting in Fig. 4c and 4d are accurate, the absolute value for applied microwave power suffers from ~20% uncertainty[26].

Due to the Klein tunneling in graphene, it is difficult to consistently obtain interdot tunnel rates below the resonator frequency[23,33]. In our device, only DQD2 can satisfy the resonance condition under typical gate voltages. DQD1 cannot be tuned into resonance because its tunnel coupling is larger than the photon energy throughout our investigated area. As a result, we can only tune the current through DQD1 by DQD2, but not the other way around, and the cross-current correlations we oberve[26] is different from the results in Ref. 9-11, where both DQDs are in resonance with the resonator.

To study two DQDs both in resonance with the resonator, a resonator with larger resonance frequency would be needed. In addition, graphene could also be replaced by GaAs[4,6], carbon nanotube[3,7], or InAs nanowire[5] systems. Though the nonlocal transport demonstrated in this T-C system may be explained by heating effects, the type of device used here may be used in future experiments entangling macroscopically-separated transport electrons, if the coherence times of the DQD qubits can be improved.

## Conclusion

Two graphene double quantum dots separated by a distance of about 60 μm are coupled to a half-wavelength reflection-line resonator. Resonator amplitude readout results show a dip near the DQD charge-degeneracy point, which can be described by the T-C model. This result demonstrates that the two distant DQDs simultaneously interact with one microwave mode, which can be valuable for the future long-distance interactions between quantum-dot-based qubits. In addition, the correlation between the currents of these two DQDs is studied. When one DQD is tuned in resonance with the resonator frequency, the DC current in the other DQD is affected and this interaction is mediated by the microwave field in the resonator. The device and the interaction demonstrated here may provide an avenue for exploring nonlocal electronic transport and correlation, although achieving a resonator-mediated coherent interaction between quantum-dot-based qubits would require quantum dots with significantly longer coherence times.

## Methods

### Device fabrication

The samples are fabricated as follows. First we mechanically exfoliated the graphene from its bulk, KISH graphite (Kyocera. Inc), to an undoped silicon chip with 285 nm oxide. In this experiment we need two pieces of few-layer graphene with proper distance between 20 to 80 μm, and we selected those that met this requirement. Second, electron beam lithography (EBL) was employed several times, starting with the fabrication of alignment marks, then plasma-etching masks and electrode patterns. The EBL resists used were PMMA 950k A4 for the first step and double-layered PMMA 950k A2 for the latter two steps. We developed the sub-micrometer patterns under 0 ℃ to establish a better control of the device specifications. Through etching out all the undesired part of the graphene sheet to realize the designed device, we strove for the all-metal-side-gated configuration as described in Ref. 22, to avoid unstable gate terminals. This etching was carried out by inductively-coupled plasma (ICP), using a 4:1 gas mixture of Oxygen to Argon. For marks and electrodes we deposited 5 nm Ti and 45 nm Au with an electron-beam evaporator. Finally, the resonator was fabricated by optical lithography followed by metal deposition in a thermal evaporator. The metal used was 200-nm-thick Al.

**Measurement setup**

The microwave response was measured using a network analyzer (NA). The input and output ports of the NA were connected to the resonator via a circulator and a 180 degree hybrid, which splits the reflected signal back to the NA. Two 30 dB attenuators were connected between the NA output port and the circulator, reducing the power applied to the resonator down to lower than -130 dBm. The reflected signal was amplified first at 4 K and then at room temperature, producing an additional gain of 60 dB, and an isolator was used to prevent noise from the amplifiers and the environment from reaching the sample. The direct transport current was amplified by a low-noise current pre-amplifier, before being measured by a digital multimeter.

**Acknowledgements**

We thank M.R. Delbecq for fruitful discussions. This work was supported by the National Fundamental Research Programme (Grant No. 2011CBA00200), the National

Natural Science Foundation (Grant Nos. 11222438, 11174267, 11274294, 61306150, 11304301, and 91121014), and the Chinese Academy of Sciences.

## Author contributions

G.W.D., D.W., S.X.L., W.C.K and H.W.J. fabricated the samples and performed the measurements. J.R.J., F.N., H.O.L., G.C., M.X. and G.C.G. provided theoretical support and analysed the data. G.P.G. supervised the project. All authors contributed to the writing of this paper.

## Competing financial interests

The authors declare no competing financial interests.

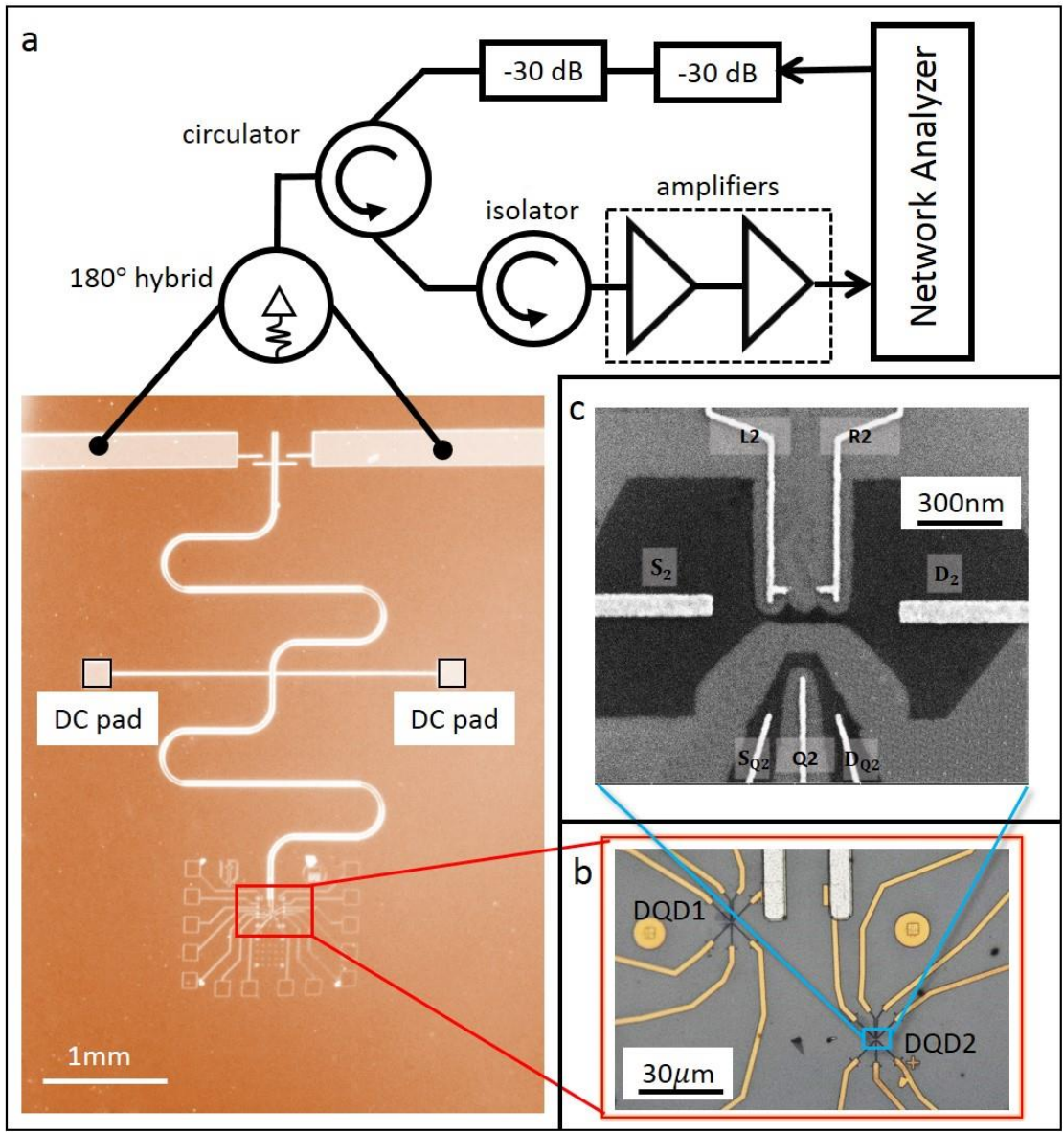

Figure 1 | Hybrid system with two graphene DQDs and a reflection-line resonator. a, Schematic and micrograph of the hybrid device. The half-wavelength reflection-line resonator is connected to the two DQDs at one end of its two striplines, while the other end is used for microwave input and output. b, The two DQDs are separated about 60 μm and are each coupled to one stripline respectively through their source leads. c, Scanning electron micrograph of a typical etched graphene DQD sample.

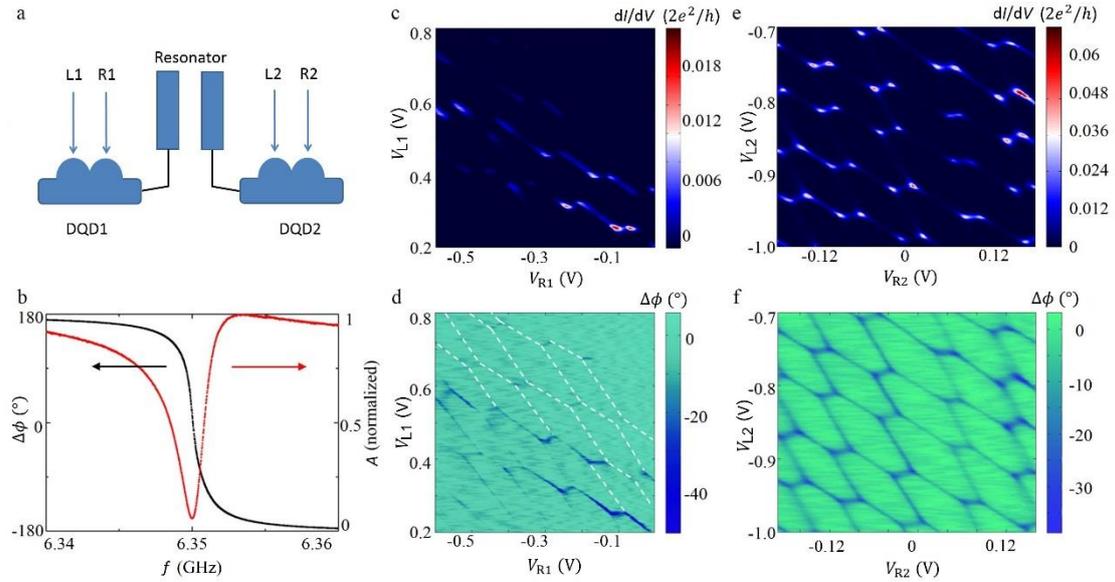

**Figure 2 | Measurement of the DQDs by transport and via the resonator. a**, Schematic diagram of the hybrid device. The electric potentials of DQD1(2) L1(2) and R1(2) gates. **b**, Spectrum of the phase (black) and amplitude (red) response of the reflection-line resonator with both DQDs in the blockade region, from which we extract the resonance frequency $f_0 = 6.35086$ GHz, internal loss $\kappa_{\text{int}}/2\pi = 0.68$ MHz, and external loss $\kappa_{\text{ext}}/2\pi = 1.32$ MHz. **c-f**, Charge-stability diagram of the two DQDs, obtained by transport measurements of the DQDs and by the response of the resonator (**c,d** for DQD1 and **e,f** for DQD2).

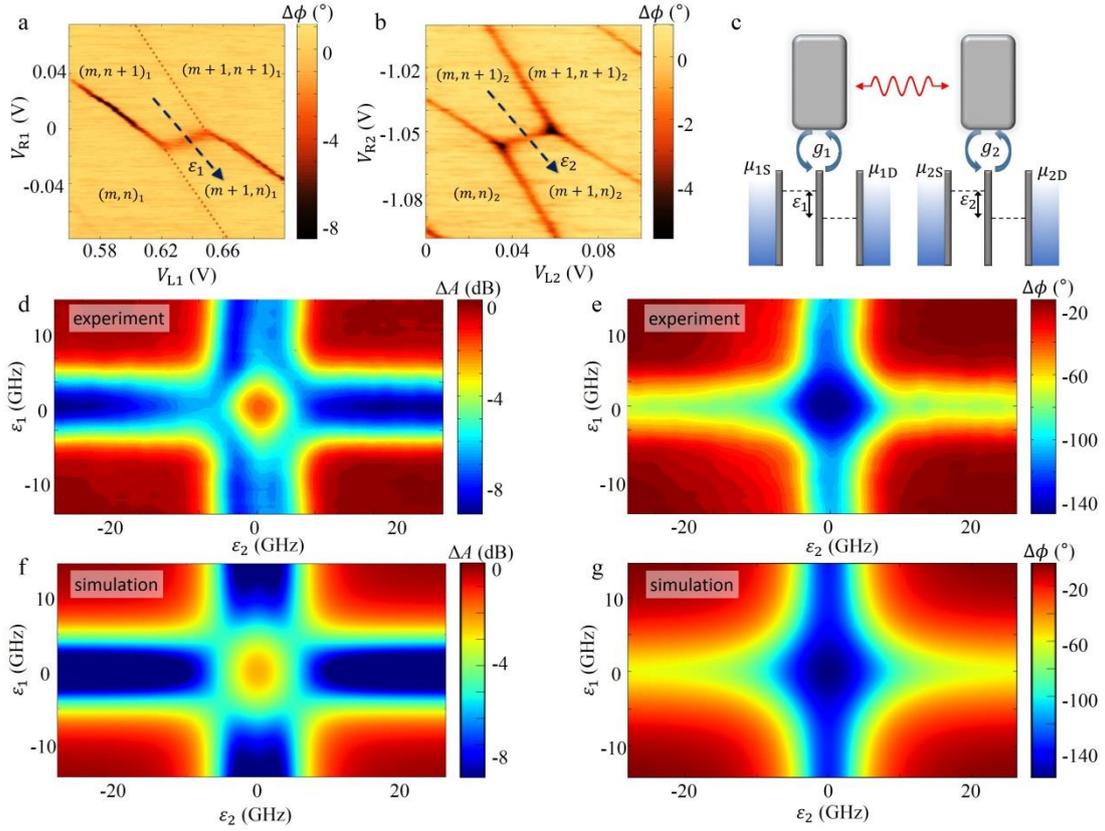

**Figure 3 | Coupling two DQDs to the resonator. a-b,** Phase response of the resonator versus gate voltage near the $(M+1, N) \leftrightarrow (M, N+1)$ charge transition for the two DQDs (**a** for DQD1 and **b** for DQD2), measured at a fixed probe frequency, $f_R = 6.35200$ GHz. The dashed arrows indicate the DQD energy detuning ($\varepsilon_1$ and $\varepsilon_2$) axes. **c**, Schematic diagram of the coupling process. The DQDs are coupled to the resonator with coupling strengths $g_1$ and $g_2$ respectively. Microwave photons are confined between two isolated superconductors, inducing an interaction between the two DQDs without direct tunneling, or capacitive coupling between them. **d-g,** Experimental (**d-e**) and simulated (**f-g**) results for the amplitude (**d** and **f**) and phase (**e** and **g**) response versus the detuning of each DQD. Parameters used in the simulation are taken from the best fits of the phase response versus detuning of each DQD[26], as denoted in **a, b**.

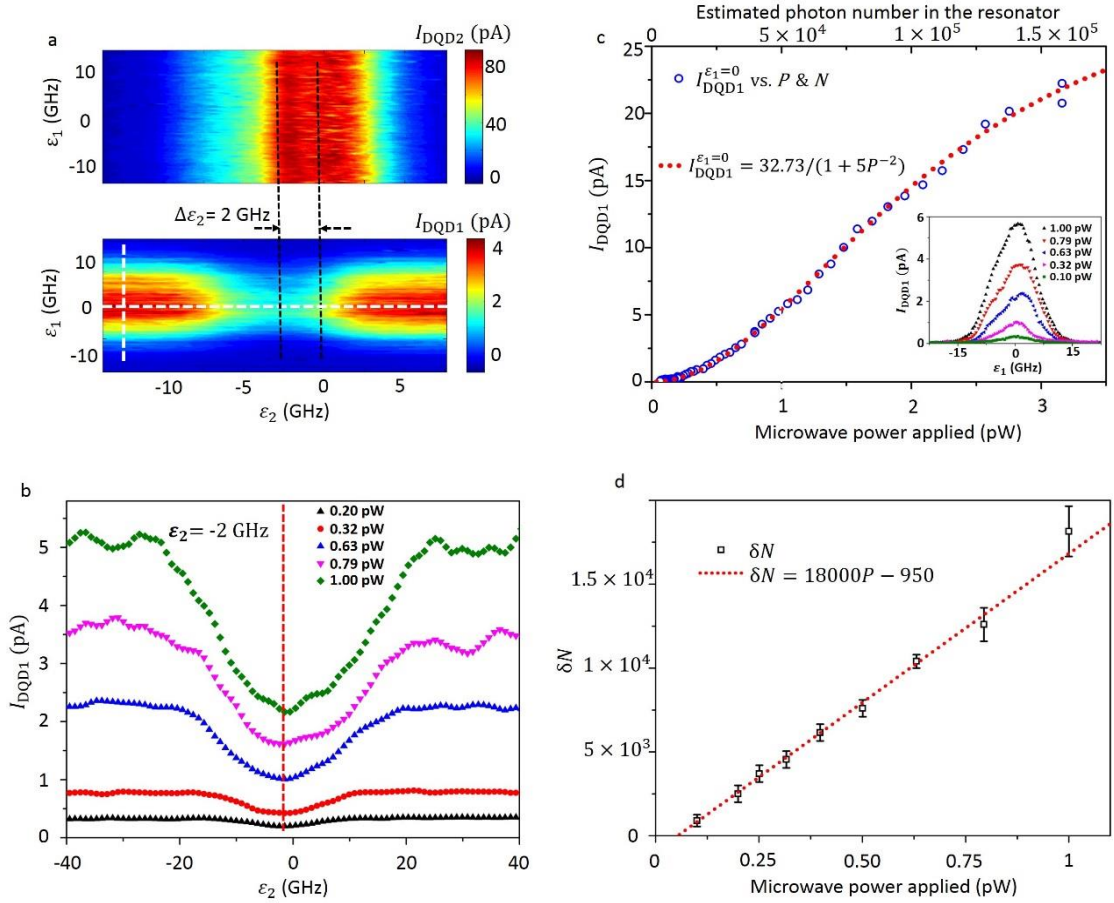

**Figure 4 | Photon-mediated current correlation for the two DQDs. a**, DC current through DQD2 and DQD1 versus the two DQDs' detuning, $\varepsilon_1$ and $\varepsilon_2$. The microwave power resonantly applied to the resonator is about 0.79 pW. **b**, $I_{DQD1}$ versus $\varepsilon_2$, with different input microwave powers. All curves can be seen as cuts at the horizontal dashed white line in **a**, fixing $\varepsilon_1 = 0$. **c**, Current peak of DQD1 versus microwave power $P$, obtained from the vertical dashed white line in **a**. Red dotted line shows the best fit of the obtained data. The inset diagram displays $I_{DQD1}$ versus $\varepsilon_1$, with different microwave powers. **d**, Photon number variation $\delta N$ due to DQD2 being on and off resonance, versus microwave applied to the resonator, obtained from the current variations shown in **b** and the photon number-current relation shown in **c**.

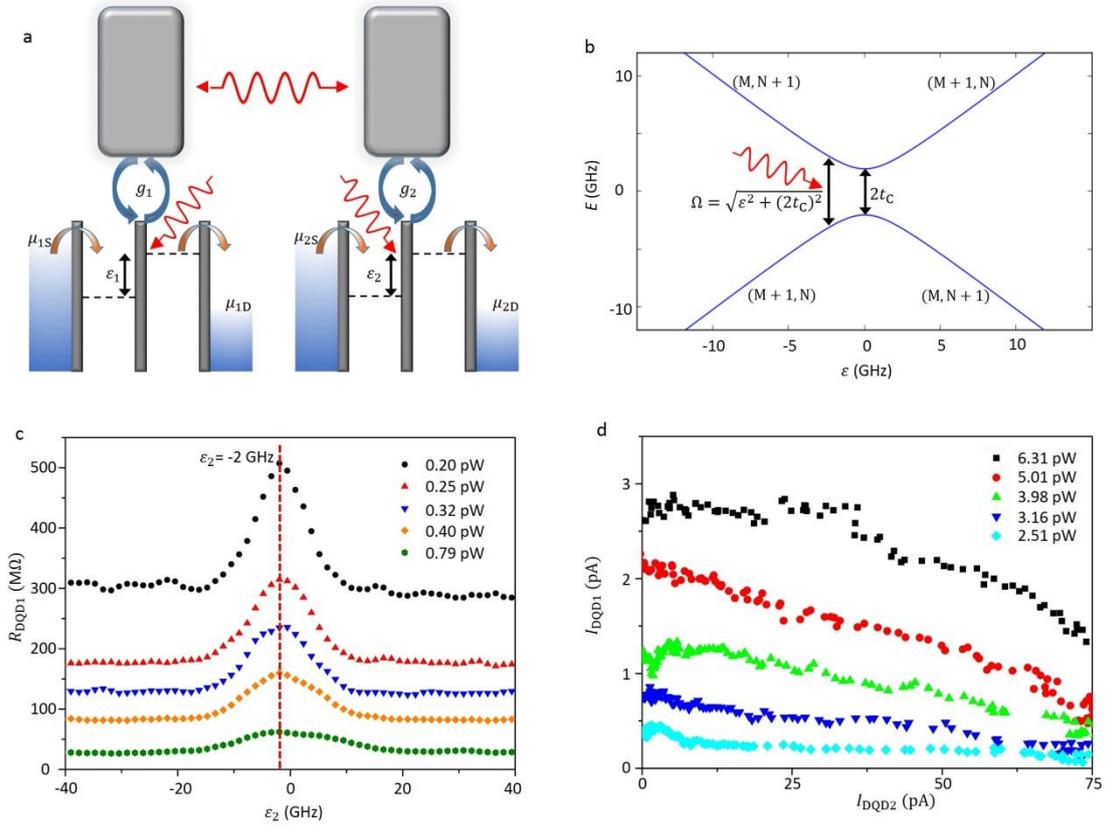

**Figure 5 | Interaction mechanism and nonlocal tunable resistance and current.**

**a,** Schematic diagram of the interaction mechanism. The two DQDs are biased at $V_{SD1} = 60\ \mu V$ and $V_{SD2} = 60\ \mu V$. When DQD2 is near resonance and $\varepsilon_2 < 0$, it strongly absorbs photons from the resonator (as shown in **b**), decreasing the microwave field in the resonator and weakening the photonic field felt by DQD1. **b,** Energy levels of a typical DQD versus detuning $\varepsilon$. **c,** Resistance of DQD1 as a function of DQD2 detuning $\varepsilon_2$, with different input microwave powers, which is obtained from Fig. 4c. **d,** Relation between $I_{DQD1}$ and $I_{DQD2}$. Here, DQD2 is tuned from the blockade region to the current peak center while DQD1 is fixed near the transition line where the current is large enough to study the current correlation.

# Supplementary materials for Coupling two distant double quantum dot to a microwave resonator

Guang-Wei Deng[1,2,*], Da Wei[1,2,*], Shu-Xiao Li[1,2], J. R. Johansson[3], Wei-Cheng Kong[1,2], Hai-Ou Li[1,2], Gang Cao[1,2], Ming Xiao[1,2], Guang-Can Guo[1,2], Franco Nori[4,5], Hong-Wen Jiang[6] and Guo-Ping Guo[1,2,+]

[1] Key Laboratory of Quantum Information, University of Science and Technology of China, Chinese Academy of Sciences, Hefei 230026, China

[2] Synergetic Innovation Center of Quantum Information and Quantum Physics, University of Science and Technology of China, Hefei, Anhui 230026, China

[3] iTHES research group, RIKEN, Wako-shi, Saitama, 351-0198 Japan

[4] CEMS, RIKEN, Wako-shi, Saitama, 351-0198 Japan

[5] Physics Department, The University of Michigan, Ann Arbor, Michigan 48109-1040, USA

[6] Department of Physics and Astronomy, University of California at Los Angeles, California 90095, USA

**Contents**

1. **Measurements of two DQDs**

2. **Capacitive coupling measurements**

3. **Tavis-Cummings model**

4. **Tunability of the energy-level splittings in SQD and DQD devices**

5. **Photon-mediated transport, photon number *N* and power *P*.**

6. **Zero-frequency cross-current correlation**

7. **Reference**

\* These authors contributed equally to this work.

+ Correspondence to: gpguo@ustc.edu.cn

# 1. Measurement of two DQDs

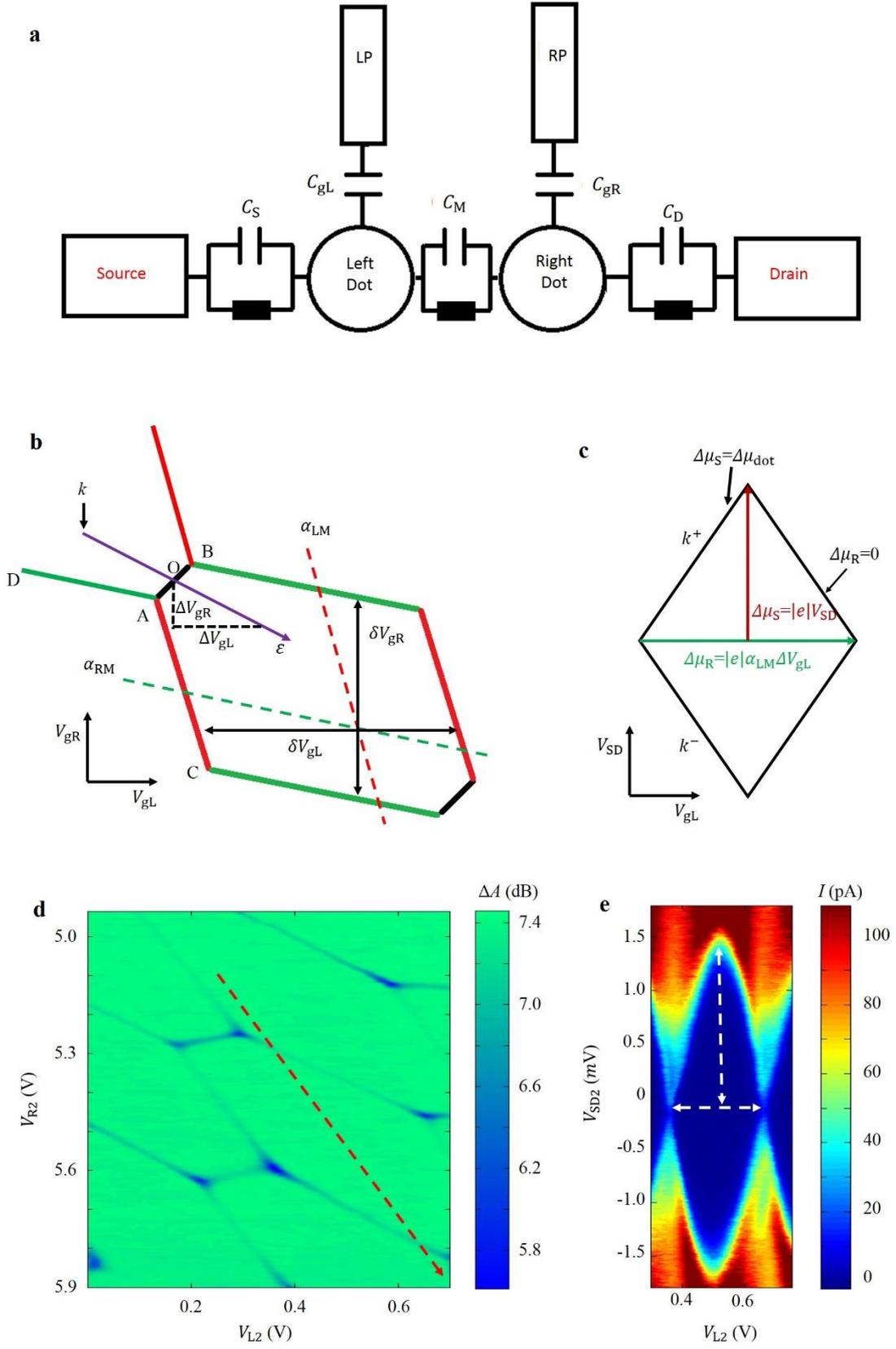

**Fig. S1: a**, Schematic diagram of a DQD. **b**, Schematic picture of the charge-stability diagram for a DQD using the constant interaction model. The red and green dashed lines are each parallel to an edge of the honeycomb, along which only one dot's energy level changes. Additionally, by applying different bias voltages, we obtain a Coulomb diamond. The purple arrow across AB indicates the sweep direction when treating the DQD as a two-level system. **c**, Schematic diagram of Coulomb diamond obtained by sweeping gate voltages along the dashed red line in **b**. Analyzing the energy shift as a function of the voltages, we can obtain $\alpha_{\text{LM}}$. **d, e,** A typical charge-stability diagram (**d**) and Coulomb diamond (**e**) of DQD2 in our device. **e** is obtained from the dashed red arrow in **d**. From the white arrows in **e**, the lever arm $\alpha_{\text{LM}}$ can be obtained.

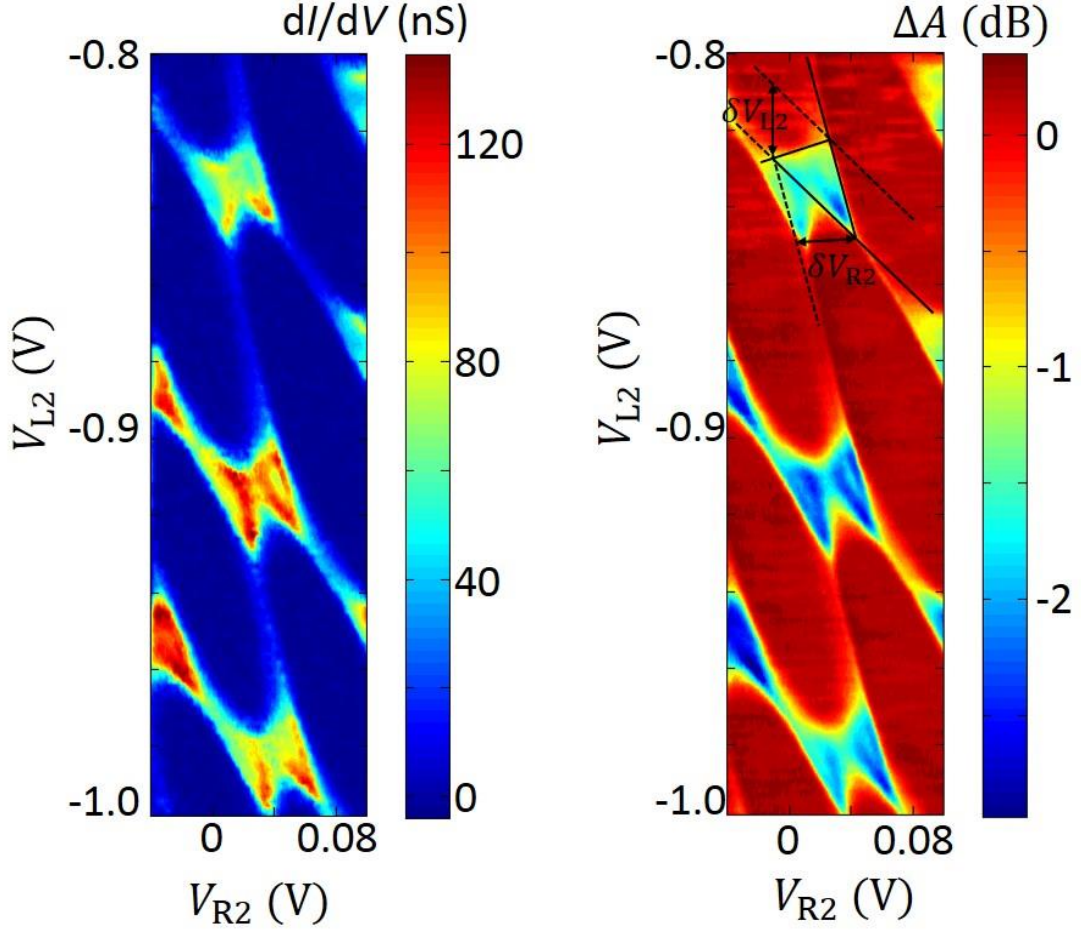

**Fig. S2:** Typical bias triangle, obtained by transport measurements (left) and the resonator signal (right). The gate lever arms can be obtained by measuring the triangle size.

First, we measure the lever arm $\alpha$ of each gate. From the constant interaction model[1], we know that when tuning the gate voltage along the red dashed line (i.e., $k_{AC} = -\frac{C_{gL}}{C_{gR}} \cdot \frac{C_R}{C_M}$), the electrochemical potential of the left dot $\mu_L$ remains unchanged while that of the right dot $\mu_R$ shifts. Therefore, we may study the charging effect of the right dot, with its gate-controlled energy written as

$$\Delta\mu_R(\Delta V_{gL}, \Delta V_{gR}) = -\frac{1}{|e|}\left(C_{gL}E_{CM}\Delta V_{gL} + C_{gR}E_{CR}k_{AC}\Delta V_{gL}\right)$$
$$= |e|\frac{C_{gL}}{C_M}\Delta V_{gL} \qquad (1)$$
$$= |e|\alpha_{LM}\Delta V_{gL},$$

where $E_{CL} = \frac{C_R}{C_L C_R - C_M^2}$, $E_{CR} = \frac{C_L}{C_L C_R - C_M^2}$ and $E_{CM} = \frac{C_M}{C_L C_R - C_M^2}$.

Additionally, by sweeping the bias voltage $V_{SD}$, a Coulomb diamond appears, giving $|e|V_{SD} = |e|\alpha_{LM}\Delta V_{gL}$ (Fig. S1b). Then we have

$$\alpha_{LM} = \frac{V_{SD}}{\Delta V_{gL}} = \left(\frac{1}{k^+} + \frac{1}{|k^-|}\right)^{-1}. \qquad (2)$$

Similarly, along the green line (i.e., $k_{AD} = -\frac{C_{gL}}{C_{gR}}\frac{C_M}{C_L}$), we obtain

$$\Delta\mu_L = |e|\frac{C_{gR}}{C_M}\Delta V_{gR} = |e|\alpha_{RM}\Delta V_{gR}. \qquad (3)$$

Using the expressions for $k_{AC}$ and $k_{AD}$, we obtain the lever arm of each gate

$$\alpha_L = \frac{C_{gL}}{C_L} = -k_{AD}\alpha_{RM}, \qquad (4)$$
$$\alpha_R = \frac{C_{gR}}{C_R} = -\frac{1}{k_{AC}}\alpha_{LM}. \qquad (5)$$

Now we derive the relation between gate voltage $\Delta V_{gL}$ and the energy detuning $\varepsilon$ during the sweep. The expressions for $\mu_L$ and $\mu_R$ are

$$\mu_L(N, M; V_{gL}, V_{gR}) = \left(N - \frac{1}{2}\right)E_{CL} + ME_{CM} - \frac{1}{|e|}\left(C_{gL}V_{gL}E_{CL} + C_{gR}V_{gR}E_{CM}\right), \qquad (6)$$

$$\mu_R(N, M; V_{gL}, V_{gR}) = NE_{CM} + \left(M - \frac{1}{2}\right)E_{CR} - \frac{1}{|e|}\left(C_{gL}V_{gL}E_{CM} + C_{gR}V_{gR}E_{CR}\right), \qquad (7)$$

taking the difference between them gives

$$\varepsilon = \varepsilon(\Delta V_{gL}, \Delta V_{gR}) = -\frac{1}{|e|}\left[C_{gL}(E_{CL} - E_{CM})\Delta V_{gL} - C_{gR}(E_{CR} - E_{CM})\Delta V_{gR}\right]$$
$$= -|e|\left[\frac{\alpha_L + k_{AD}\alpha_R}{1 - \frac{k_{AD}}{k_{AC}}} - \frac{\alpha_R + \frac{1}{k_{AC}}\alpha_L}{1 - \frac{k_{AD}}{k_{AC}}}k\right]\Delta V_{gL}. \qquad (8)$$

Here $k$ is the slope in the gate voltage sweep process (Fig. S1b). Substituting $\alpha_L$ and

$\alpha_R$ with parameters in Fig. S1b yields

$$\alpha_L + k_{AD}\alpha_R = \frac{C_{gL}}{C_L} - \frac{C_{gL}}{C_{gR}}\frac{C_M}{C_L}\frac{C_{gR}}{C_R} = \frac{C_{gL}}{C_L}\frac{C_R - C_M}{C_R}, \tag{9}$$

$$\alpha_R + \frac{1}{k_{AC}}\alpha_L = \frac{C_{gR}}{C_R}\frac{C_L - C_M}{C_L}. \tag{10}$$

With $k_{AB} = \frac{C_{gL}}{C_{gR}}\frac{C_R - C_M}{C_L - C_M}$ (the slope from point A to B in Fig. S1b), and $C_{gL(R)} = \frac{|e|}{\delta V_{gL(R)}}$, finally we have

$$\varepsilon = -|e|\Delta V_{gL}\frac{\frac{C_{gL}}{C_L}\cdot\left(1 - \frac{C_M}{C_R}\right)}{1 - \frac{k_{AD}}{k_{AC}}}\cdot\left(1 - \frac{k}{k_{AB}}\right)$$

$$= -|e|\alpha_L \cdot \frac{1 - \frac{k}{k_{AB}}}{1 - \frac{k_{AD}}{k_{AC}}}\cdot\left(1 + \frac{1}{k_{AC}}\cdot\frac{\delta V_{gR}}{\delta V_{gL}}\right)\cdot \Delta V_{gL}. \tag{11}$$

## 2. Capacitive coupling measurement

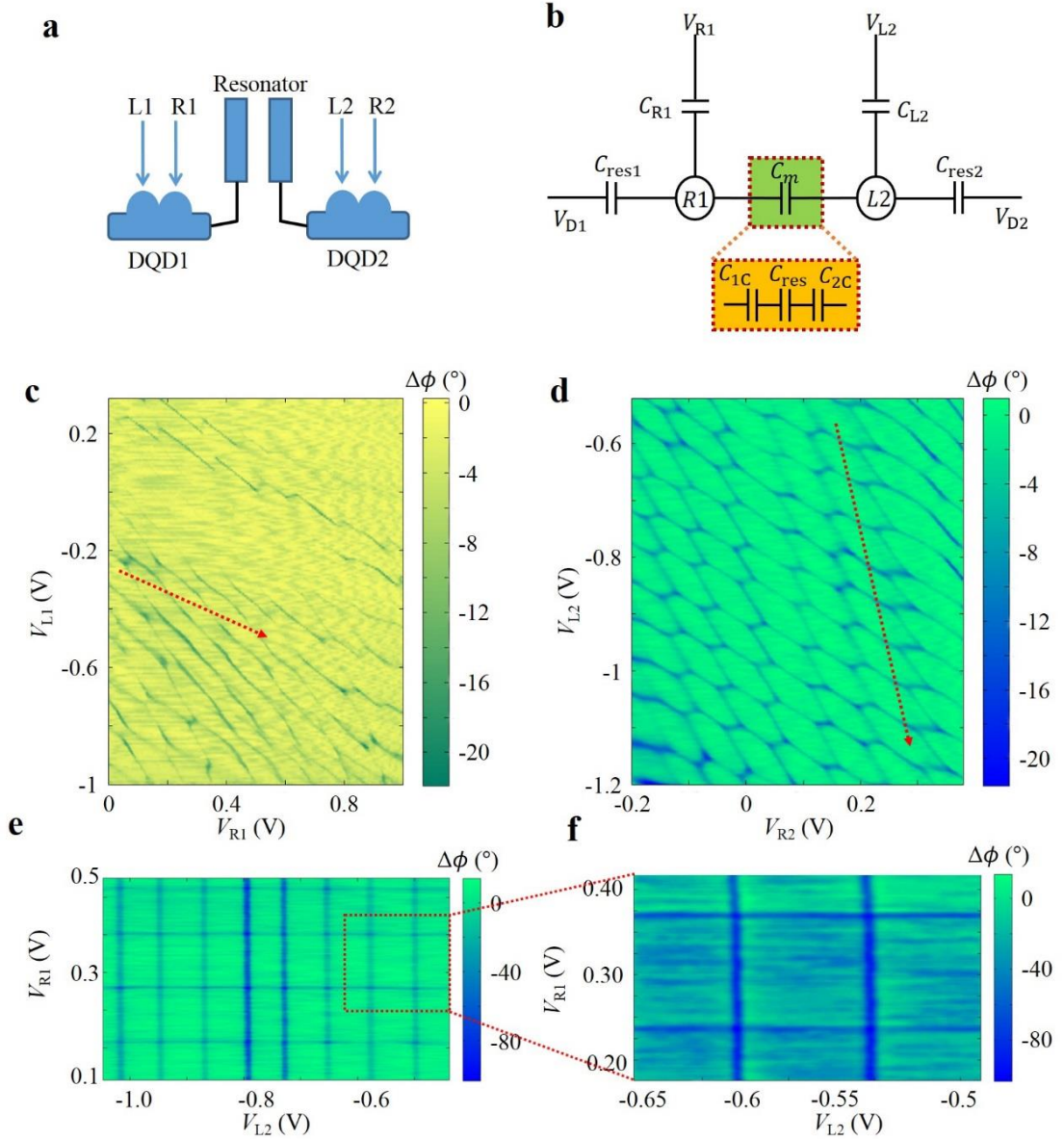

**Fig. S3: a**, Schematic diagram of our sample. **b**, Model of the effective DQD formed by R1 and L2, when tuning the gate voltages as shown in **c** and **d**. Here $C_{iC}$ is the coupling capacitance between the $i$th DQD and the resonator. $C_{\text{res}}$ is the capacitance between the two parallel reflection lines of the resonator. **c**, Charge-stability diagram of DQD1. Along the red arrow, the charge number of the L1 dot does not change, and the signal reflects the charging effect of the R1 dot. **d**, Charge-stability diagram of DQD2. **e.f.** Charge-stability diagram of the R1-L2 DQD, obtained by sweeping along red dashed arrows in **c, d**. In the constant interaction model, such rectangular grids indicate a near-zero middle capacitance $C_m$ and capacitive interaction between the two DQDs is therefore negligible.

Figure. S3a,b show the schematic circuit diagrams of the hybrid device. For simplicity, we consider an effective DQD formed by the R1 dot of DQD1 and the L2 dot of DQD2 (Fig. S3b). To obtain the charge-stability diagram of this effective DQD, we sweep the gate voltages along the red dashed arrows denoted in Fig. S3c, d. Joint readout results (Fig. S3e) suggest this effective DQD has negligible interdot capacitive coupling[1], i.e., $C_\mathrm{m} = 0$, as the single dot charging lines cross each other perpendicularly.

### 3. Tavis-Cummings model

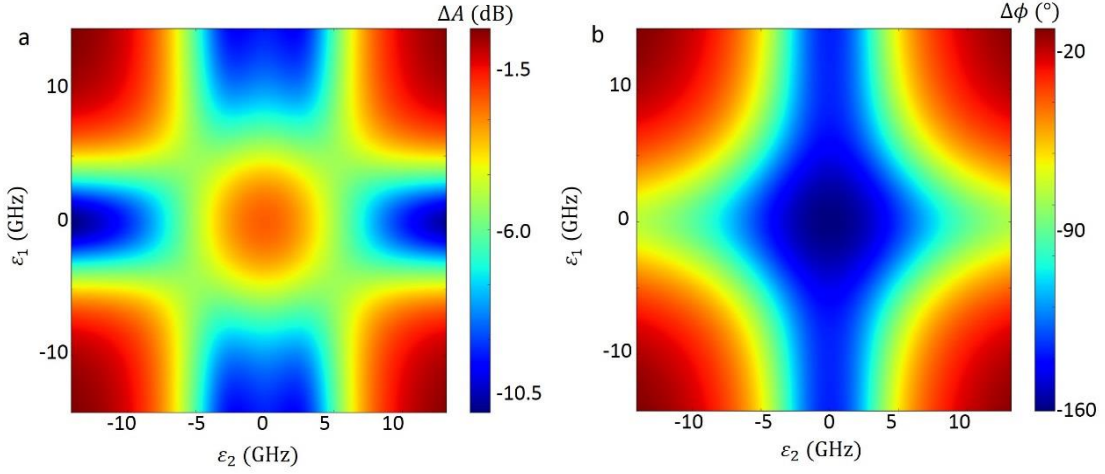

**FIG. S4: Simulation results. a**, A simulation of the weak-coupling regime, which is similar to our experimental results. The DQD parameters are $g_{C1} = 40$ MHz, $g_{C2} = 29$ MHz, $2t_{C1} = 7.2$ GHz, $2t_{C2} = 7.2$ GHz, $\Gamma_1 = (\frac{\gamma_1}{2} + \gamma_2)_{DQD1} = 3.6$ GHz, $\Gamma_2 = (\frac{\gamma_1}{2} + \gamma_2)_{DQD2} = 3.0$ GHz, where $g_{C1}$ ($g_{C2}$) $< \Gamma_1$ ($\Gamma_2$).

In the joint-readout experiment, our sample can be seen as two quantum two-level systems that are dipole coupled to a resonator, and it can therefore be described by the Tavis-Cummings Hamiltonian[3]

$$H = \omega_0 a^\dagger a + \sum_{i=1,2} [\frac{1}{2}\Omega_i \sigma_{zi} + g_i(\sigma_{i+}a + \sigma_{i-}a^\dagger)], \tag{12}$$

where $g_i = g_{Ci}\frac{2t_{Ci}}{\Omega_i}$ and $\Omega_i = \sqrt{(2t_{Ci})^2 + \varepsilon_i^2}$. $\omega_0$ is the resonance frequency of resonator, $\varepsilon_i$, $t_{Ci}$ and $g_{Ci}$ denote the detuning, the tunneling matrix element and the DQD-resonator coupling constant of DQD$i$, respectively.

To determine the reflected signal, using input-output theory[4], we write down the Heisenberg-Langevin equations of motion for the operators $a$ and $\sigma_{i-}$

$$\dot{a}(t) = -j\omega_0 a(t) - j\sum_{i=1,2} g_i \sigma_{i-} - \frac{1}{2}(\kappa_e + \kappa_i)a(t) + \sqrt{\kappa_e} a_{in}(t), \tag{13}$$

$$\dot{\sigma}_{i-}(t) = -j\Omega_i \sigma_{i-}(t) + jg_i a(t)\sigma_{zi}(t) - \frac{1}{2}\gamma_{1i}\sigma_{i-}(t) - \gamma_{2i}\sigma_{i-}(t), \tag{14}$$

where $\kappa_e$ ($\kappa_i$) is the external (internal) dissipation rate of the resonator. In what follows, we assume that the quantum dot stays in its ground state, leading to the replacement $\sigma_{zi} \to -1$. Fourier transformation of the remaining linear equations then gives

$$-j\omega \sigma_{i-}(\omega) = -j\Omega_i \sigma_{i-}(\omega) - jg_i a(\omega) - \frac{1}{2}\gamma_{1i}\sigma_{i-}(\omega) - \gamma_{2i}\sigma_{i-}(\omega), \tag{15}$$

$$-j\omega a(\omega) = -j\omega_0 a(\omega) - j\sum_{i=1,2} g_i \sigma_{i-}(\omega) - \frac{1}{2}(\kappa_e + \kappa_i)a(\omega) + \sqrt{\kappa_e} a_{in}(\omega). \tag{16}$$

Using the boundary condition $a_{in} + a_{out} = \sqrt{\kappa_e} a$, combined with the results above, we obtain

$$\sigma_{i-}(\omega) = -\frac{jg_i}{j(\Omega_i - \omega) + \frac{1}{2}\gamma_{1i} + \gamma_{2i}} a(\omega) = -j\chi_i a(\omega), \tag{17}$$

$$\left[j(\omega_0 - \omega) + g_1\chi_1 + g_2\chi_2 + \frac{1}{2}(\kappa_e + \kappa_i)\right] a(\omega) = \sqrt{\kappa_e} a_{in}(\omega). \tag{18}$$

Finally, we obtain the input-output relation

$$S_{11} = \frac{a_{out}}{a_{in}} = -\frac{j(\omega_0 - \omega) + g_1\chi_1 + g_2\chi_2 + \frac{1}{2}(\kappa_i - \kappa_e)}{j(\omega_0 - \omega) + g_1\chi_1 + g_2\chi_2 + \frac{1}{2}(\kappa_i + \kappa_e)}, \tag{19}$$

$$\chi_i = \frac{g_i}{j(\Omega_i - \omega) + \frac{1}{2}\gamma_{1i} + \gamma_{2i}}, \quad i = 1,2. \tag{20}$$

To compare this to our experimental data, instead of $S_{11}$, we define the amplitude $A = 20 \times \lg|S_{11}|$ and the argument $\phi = \arg(S_{11})$ of $S_{11}$. Here $A$ is in dB unit and $\phi$ is in degree unit. Both of them can be directly measured by a network analyzer. With the parameters of each DQD obtained by independent experiments following the method in Ref. 5, we can reproduce the joint readout results in Fig. 3 in the main text. In this case, the DQD parameters used are $g_{C1} = 40$ MHz, $g_{C2} = 29$ MHz, $2t_{C1} = 7.2$ GHz, $2t_{C2} = 7.2$ GHz, $\Gamma_1 = (\frac{\gamma_1}{2} + \gamma_2)_{DQD1} = 3.6$ GHz, $\Gamma_2 = (\frac{\gamma_1}{2} + \gamma_2)_{DQD2} = 3.0$ GHz, and the resonator parameters are $\kappa_i/2\pi = 0.684$ MHz, $\kappa_e/2\pi = 1.318$ MHz, $\omega_0/2\pi = 6.35086$ GHz. Figure S4 shows two typical simulation results for amplitude response (a) and phase response (b) of the resonator that are representative for the weak-coupling regime. However, due to the large dephasing rates in DQD systems, we cannot reach the strong-coupling regime in our device.

From equation (19), it is clear that the effect of the two DQDs on the resonator signal $S_{11}$ is the term $g_1\chi_1 + g_2\chi_2$, where $\chi_i$ is the susceptibility[6] of the $i$th DQD. If we write $g_1\chi_1 = \delta\kappa_1 + j\delta\omega_1$, $g_2\chi_2 = \delta\kappa_2 + j\delta\omega_2$, and define $\delta\omega = \delta\omega_1 + \delta\omega_2$, $\delta\kappa = \delta\kappa_1 + \delta\kappa_2$, then equation (19) can be rewritten as

$$S_{11} = -\frac{j[(\omega_0 + \delta\omega) - \omega] + \frac{1}{2}[(2\delta\kappa + \kappa_i) - \kappa_e]}{j[(\omega_0 + \delta\omega) - \omega] + \frac{1}{2}[(2\delta\kappa + \kappa_i) + \kappa_e]}. \quad (21)$$

Here $\delta\omega$ is the resonator frequency shift due to the DQDs, and $2\delta\kappa$ is the change of the internal resonator decay rate due to the DQDs, which produces a broadening of the linewidth. Notice that $g_i\chi_i = 0$ when $\varepsilon_i \to \infty$, leading to a pure resonator response.

Now we define the contribution of the DQDs to the signal using

$$\Delta A(\varepsilon_1, \varepsilon_2) = A(\varepsilon_1, \varepsilon_2) - A(\infty, \infty) = \Delta A(\delta\omega_1, \delta\omega_2; 2\kappa_1, 2\kappa_2) = \Delta A(\delta\omega; \delta\kappa) \quad (22)$$

$$\Delta\phi(\varepsilon_1, \varepsilon_2) = \phi(\varepsilon_1, \varepsilon_2) - \phi(\infty, \infty) = \Delta\phi(\delta\omega_1, \delta\omega_2; 2\kappa_1, 2\kappa_2) = \Delta\phi(\delta\omega; \delta\kappa) \quad (23)$$

The $\varepsilon_i$-induced frequency shift and internal decay increase can be written as $\delta\omega_1 +$

$\delta\omega_2$ and $2\kappa_1 + 2\kappa_2$, respectively (see Fig. S5). However, since $S_{11}$ is nonlinear in these variables, $\Delta A$ and $\Delta\phi$ are not additive, i.e., $\Delta A(\varepsilon_1, \varepsilon_2) \neq \Delta A(\varepsilon_1, \infty) + \Delta A(\infty, \varepsilon_2)$, $\Delta\phi(\varepsilon_1, \varepsilon_2) \neq \Delta\phi(\varepsilon_1, \infty) + \Delta\phi(\infty, \varepsilon_2)$. This phenomenon is observed in our joint readout experiment, and the results are explained by the T-C model, as shown in the main text (see Fig. 3).

In the T-C model, there is a coherent coupling between the DQDs mediated by the resonator since they both exchange real or virtual photons with the resonator. This is in contrast to the results in Fig. S3. In Fig. S3, $\Delta A$ and $\Delta\phi$ sum directly. At the cross point, as the charging energy is much larger than photon energy, the signal indicates a quantum admittance, which is a linear response[7]. This kind of direct summation in $\Delta A$ and $\Delta\phi$ leads to a direct-crossing picture (Fig. S3e,f). However, the T-C model show a non-linear relation in the $\Delta A$ diagram (Fig. S4a).

There may be higher-order processes in this kind of hybrid system, especially when two DQDs are source-drain biased, and these processes could contribute to the current and low-frequency noise in both DQDs[8-10]. In our joint readout measurements (see Fig. 3), all leads were grounded in order to avoid higher-order processes, as such effects are not included in the T-C Hamiltonian.

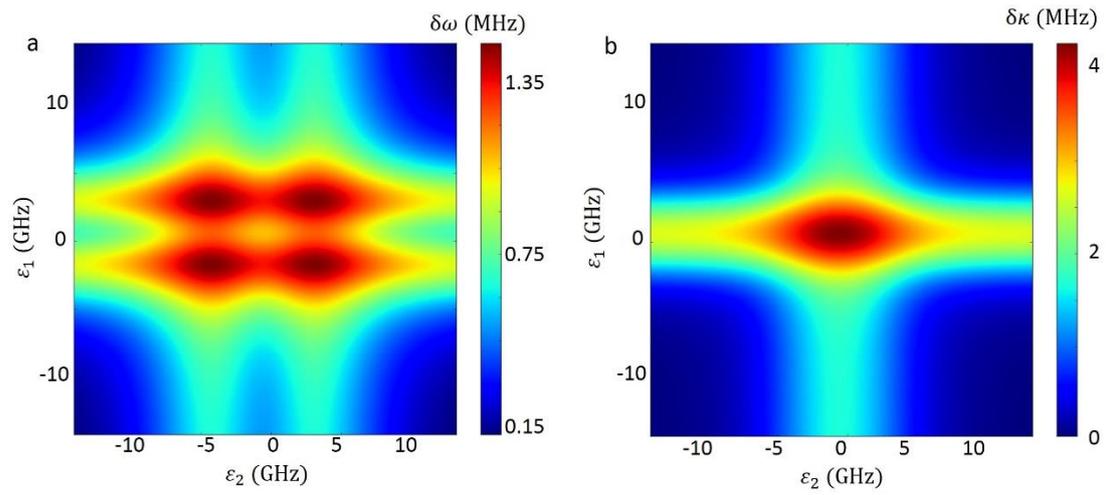

**FIG. S5: Simulation results. a**, Frequency shift as a sum of the shift due to each DQD. **b**, Internal decay rate increase.

## 4. Tunability of the energy level-splittings in SQD and DQD devices

In this section we analyze and compare the tunability of the energy-level splittings in SQD and DQD devices. In QD devices, there are two types of discrete energy levels; the charging energy and the single particle energy levels. The former stems from the physics of charged particles confined in a box, interacting through the Coulomb interaction. The level spacing, usually referred to as "charging energy", denoted as $E_c$, is the energy needed to add one more electron into the QD and its value primarily depends on the size of the QD. Typical values of $E_c$ range from 1 to 100 meV[1,11], which is far larger than the energy scale of the resonator photons (~30 μeV).

Single particle energy levels are due to internal degrees of freedom, such as orbit and spin. For spin degrees of freedom, inducing and controlling the energy-level splitting requires application of an external magnetic field. Note that Al ceases to be superconducting when the external field exceeds 300 mT, and this sets an upper limit on the obtainable Zeeman splitting of the spin states. In addition, applying a magnetic field not only changes the desired energy splitting, it would likely also change other properties of the sample dramatically.

For orbital degrees of freedom (also often be referred to as charge states), the energy splitting ranges from ~10 μeV to ~1 meV. Admittedly, it covers our desired range. However, gate-potential-defined SQD devices are usually implemented with plunger gates and defining gates. Tuning the plunger-gate potential shifts the energy levels in the QD as a whole and leaves the energy-level splitting unchanged. For the defining gates, the changes in the defining potential change both the shape and the size of the QD. Experimentally, it requires great effort to tune the energy splitting without changing other fundamental properties, including the QD shape and the barrier tunneling rates. Regarding the shape-defined SQDs, such as carbon nanotubes and etched graphene nanoribbons, control over the single particle energy level splitting is even more difficult.

The DQDs are formed either by their shape or applied potentials. Still, near a transition line in the charge-stability diagram, the energy splitting between the charge states in the left and right dot is $\Omega = \sqrt{\varepsilon^2 + (2t_\text{C})^2}$. This splitting can be directly controlled by gate voltages. Note that $2t_\text{C}$ typically ranges from ~1 to ~100 μeV.

## 5. Photon-mediated transport, photon number *N* and power *P*

In this section we elaborate on the experimental details related to the section *"Photon-mediated electron transport"* in the main text. Experimentally, the microwave power is generated by the network analyzer (NA). Before this power reaches the resonator, it is attenuated by 30 dB and 36 dB attenuators, acting consecutively. Furthermore, cables, connectors and the wire bonding to the silicon chip contribute to a total estimated attenuation of about 9 dB ($\pm$ 3 dB). All these add up to a total attenuation of 75 dB.

The power that is absorbed by the resonator, *P*, establishes equilibrium in the resonator, and can be converted to a photonic field defined by the average number of photon inside the resonator, *N*, where $N = 4P\kappa_e/(\hbar\omega_0(\kappa_i + \kappa_e)^2)$. The factor of conversion is determined by the rate of dissipation of the resonator[2,12,13].

When DQD2 is far off resonance, DQD1's peak current reflects the field strength in the resonator. Thus we start from the empirical law $I_{\text{DQD1}}^{\varepsilon_1=0} = 32.73/(1 + 5P^{-2})$. In the language of microwave power, our experimental results can be also explained as follows: the power that reaches the resonator is dissipated through DQD1, DQD2 and other channels to the environment. In equilibrium, the total dissipation equals to *P*. When DQD2 is in resonance ($\Omega_2 \sim hf_0$), more microwave power is dissipated through DQD2 and less through DQD1. Therefore, DQD2 works as an attenuator for DQD1. The empirical law can now be understood as the relation between the peak conductance of DQD1 and the power dissipated through it. If we denote the latter value as $P_{\text{DQD1}}$ and write it as α*P* (α<1), the empirical law becomes $I_{\text{DQD1}}^{\varepsilon_1=0} =$ $32.73/(1 + 5\alpha^2 P_{\text{DQD1}}^{-2})$ and differs only in the coefficient. Note that α is an experimental constant denoting DQD1's contribution to the total dissipation when DQD2 is off resonance.

Quantitatively, if we consider $\delta P_{\text{DQD1}}$ (the difference in power dissipated through DQD1 when DQD2 is in and off resonance) versus *P*, the relation is $\delta P_{\text{DQD1}} = 0.36\alpha P + 0.019\alpha$ instead of $\delta N = 18000P - 950$. The linearity in $\delta P_{\text{DQD1}}$ versus *P* suggests that, within the power range from 0.1 to 1.0 pW, the power dissipation through DQD1 decreases by a constant factor, when DQD2 is tuned from off resonance to in resonance.

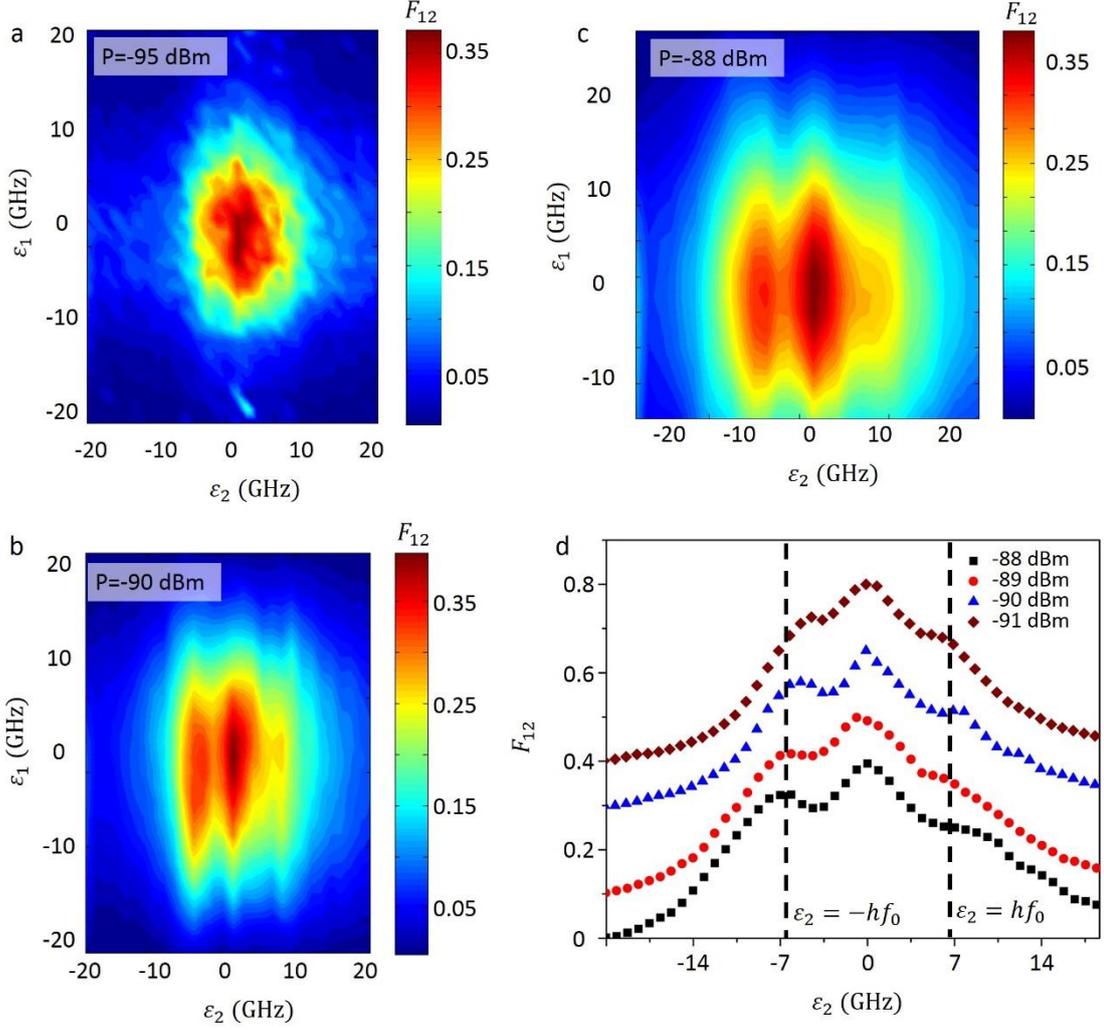

**FIG. S6: Zero-frequency cross-current correlation of the two DQDs. a-c**, Zero-frequency cross-current correlation $F_{12} = S_{12}/\sqrt{I_{DQD1} I_{DQD2}}$ as a function of the detuning of each DQD, with different input microwave powers. **d,** $F_{12}$ versus $\varepsilon_2$, with different input microwave powers. Here $\varepsilon_1 = 0$ is fixed and each curve is offset by 0.15 from the previous for clarity. All curves have a triple-peak structure, with peaks at $\varepsilon_2 = -hf_0$, $\varepsilon_2 = 0$, and $\varepsilon_2 = hf_0$.

## 6. Zero- frequency cross-current correlation

Ref. 8-10 predict an enhanced zero-frequency cross-current correlation $F_{12} = S_{12}/\sqrt{I_{\text{DQD1}}I_{\text{DQD2}}}$ (Fano factor), when both DQDs are tuned in resonance with the resonator. Here $S_{12}$ is the zero-frequency correlation between the currents through DQD1 and DQD2, which can be directly obtained with a two-channel dynamic signal analyzer (Stanford Research Systems Model SR785). Although only DQD2 can be tuned in resonance in our device, we still observe an enhanced $F_{12}$ due to the DQD-resonator interaction (see Fig. S6). For small input microwave power, the Fano factor shows a peak around $\varepsilon_1 = 0$ and $\varepsilon_2 = 0$ (Fig. S6a), while a triple-peak structure appears at larger input power (Fig. S6b,c). The peak positions are around $\varepsilon_2 = -hf_0$, $\varepsilon_2 = 0$, and $\varepsilon_2 = hf_0$ (Fig. S6d). Although the mechanism is not clear to us, these results show a clear correlation between the two currents, induced by the interaction between the DQDs and the resonator.